\documentstyle[12pt,fleqn]{article}
\catcode`\@=11
\@addtoreset{equation}{section}

\def\be{\begin{equation}}
\def\ee{\end{equation}}
\def\ba{\begin{eqnarray}}
\def\ea{\end{eqnarray}}

\def\la{\mathrel{\mathpalette\fun <}}
\def\ga{\mathrel{\mathpalette\fun >}}
\def\fun#1#2{\lower3.6pt\vbox{\baselineskip0pt\lineskip.9pt
        \ialign{$\mathsurround=0pt#1\hfill##\hfil$\crcr#2\crcr\sim\crcr}}}

\begin{document}
\begin{titlepage}
\null\vspace{-70pt}
\begin{flushright}FERMILAB-Pub-95/013-A\\
January, 1995 \\
revised version: August, 1995
\end{flushright}

\vspace{.2in}
\baselineskip 24pt
\centerline{\large \bf{Simple analytical methods for computing the
gravity-wave}}
\centerline{\large \bf{contribution to the cosmic background radiation
anisotropy }}

\vspace{.3in}
\centerline{Yun Wang}
\vspace{.2in}
\centerline{{\it NASA/Fermilab Astrophysics Center}}
\centerline{\it Fermi National Accelerator Laboratory, Batavia, IL 60510-0500}

\vspace{.4in}
\centerline{\bf Abstract}
\begin{quotation}

We present two simple analytical methods for computing the
gravity wave contribution to the cosmic background radiation (CBR)
anisotropy in inflationary models; one method uses a time-dependent
transfer function, the other method uses an approximate gravity-wave
mode function which is a simple combination of the lowest order
spherical Bessel functions. We compare the CBR anisotropy tensor
multipole spectrum computed using our methods with the previous result
of the highly accurate numerical method, the ``Boltzmann'' method.
Our time-dependent transfer function is more accurate than the
time-independent transfer function found by Turner, White, and Lidsey;
however, we find that the transfer function method is only good
for $l \la 120$. Using our approximate gravity-wave mode function,
we obtain much better accuracy; the tensor multipole spectrum we find
differs by less than 2\% for $l \la 50$, less than 10\% for $l \la 120$,
and less than 20\% for $l \leq 300$ from the ``Boltzmann'' result.
Our approximate graviton mode function should be quite useful
in studying tensor perturbations from inflationary models.

\end{quotation}

PACS index numbers: 04.30.+x, 98.80.Cq, 98.80.Es

\end{titlepage}

\baselineskip=24pt

\section{1. Introduction}

The cosmic background radiation (CBR) anisotropy
places stringent constraints on theories of the early Universe.
Among these theories, the best studied are the inflationary models,
which are strongly motivated because they solve the famous problems
(the flatness problem, the smoothness problem, the structure formation
problem) of the standard cosmology.
Tensor (gravity-wave) and scalar (density) metric perturbations
are generated in the very early Universe due to quantum fluctuations
arising during inflation. Both tensor and scalar perturbations
contribute to anisotropy in the temperature of the CBR.
While the scalar contribution to the CBR anisotropy involves more
complicated physics, the tensor contribution to the CBR anisotropy arises
only through the Sachs-Wolfe effect [\ref{SW}] as follows.
As photons of the CBR propagate toward us from the last scattering
surface, their paths are perturbed by the metric perturbations
due to the primordial gravity waves.
The perturbed energies of these photons result in temperature
fluctuations in the sky that we observe. The CBR temperature
fluctuation is conventionally expanded into spherical harmonics:
\be
\frac{\delta T}{T}(\theta, \phi)=\sum_{lm} \, a_{lm}Y_{lm}(\theta, \phi)
\ee
In this paper, we present two simple analytic methods of
computing the tensor contribution to the variance in the
CBR temperature multipole moments, $\langle |a_{lm}|^2 \rangle $,
one method makes use of a time-dependent transfer function,
the other uses an approximate gravity-wave
mode function which is a simple combination of the lowest order
spherical Bessel functions.
Our methods provide adequate accuracy for normalizing the tensor
perturbations arising from inflationary theories to the
observable CBR anisotropy. Our method using the approximate gravity-wave
mode function is accurate enough to be used in studying the CBR anisotropy
tensor multipole spectrum to large $l$ ($l \leq 300$).

We use the results by Dodelson, Knox, and Kolb [\ref{DKK}] as the
standard for comparison. They considered a Universe with both matter
and radiation, and used numerical methods to evolve the photon
distribution function using first-order perturbation theory
of the general relativistic Boltzmann equation for radiative transfer.
Their results (which we refer to as ``Boltzmann'') should be equivalent
to the results obtained using the Sachs-Wolfe formula (see below).
The ``Boltzmann'' method has no simple analytical formulation,
and the exact result using the Sachs-Wolfe formula involves complicated
spheroidal wavefunctions [\ref{KA}]; hence it is of great interest
to find a simple transfer function, or a simple approximate graviton
mode function, which can be used analytically to
compute the CBR anisotropy tensor multipole moments to
sufficient accuracy.

Our first method has been motivated by the tensor transfer function found
by Turner, White, and Lidsey (TWL) [\ref{TWL}]. Their transfer function takes
into account the effect of the Universe becoming matter-dominated gradually.
The tensor multipole moments obtained using their transfer
function, however, differ from the ``Boltzmann'' results by up to over
30\% for $l <100$. The reason for this substantial discrepancy
is that their transfer function contains no time evolution,
although they did give the expression for the time-dependence of
the transfer function for short wavelength modes.
We modify their transfer function by
accounting for the difference in time evolution between long and short
wavelength modes.

Our second method has been motivated by intuition. Since the gravity-wave
mode function is analytically known for both matter and radiation dominated
eras, it should be possible to construct a simple approximate gravity-wave
mode function by smooth interpolation.
As expected, the resultant mode function is much closer to the true
mode function than the much used matter-dominated mode function.
When used in computing the CBR anisotropy tensor spectrum,
the approximate gravity-wave mode function gives very accurate results.

\section{2. Gravity wave contribution to CBR anisotropy}
The rms temperature fluctuation averaged over the sky for a given
experiment is given by
\be
\langle \left(\frac{\delta T}{T}\right)^2 \rangle
=\sum_{l\geq 2} \frac{2l+1}{4\pi} \langle |a_{lm}|^2 \rangle W_l,
\ee
where $W_l$ is the appropriate response function for the experiment.
For an experiment with two antennas of Gaussian beam width $\sigma$ separated
by angle $\theta$, the temperature difference between the two antennas
is measured; the response function is
\be
W_l= 2[1-P_l(\cos\theta)] \, e^{-(l+1/2)^2 \sigma^2}.
\ee
We have followed the notation of Ref.[\ref{TWL}].

Tensor perturbations generated by inflation are stochastic in nature
[\ref{KT}].
Let us expand the gravity-wave perturbation in plane waves
\be
h_{jk}({\bf x}, \tau)=(2\pi)^{-3} \int {\rm d}^3k\, h^i_{\bf k}(\tau)
\epsilon^i_{jk}\, e^{-i{\bf k\cdot x}},
\ee
where $\epsilon^i_{jk}$ is the polarization tensor and $i=\times$, $+$
in the transverse traceless gauge (in which $h_{00}=h_{0j}=0$).
We have
\be
h^i_{\bf k}(0)=A(k) a^i({\bf k}),
\ee
where $a^i({\bf k})$ is a random variable with statistical expectation value
\be
\langle a_i({\bf k})\, a_j({\bf q}) \rangle =
k^{-3}\delta^{(3)}({\bf k}-{\bf q})
\, \delta_{ij},
\ee
and the spectrum of gravity waves generated by inflation is
\be
\label{eq:A(k)}
A^2(k)=\frac{H^2}{\pi^2 M_{\rm PL}^2}=\frac{8}{3\pi} \frac{V}{M_{\rm PL}^4},
\ee
where $V$ is the value of the inflaton potential when the mode with
comoving wavenumber $k$ crosses outside the horizon during inflation.

The tensor contribution to the variance of the multipoles is given by
\be
\label{eq:|a_lm|^2}
\langle |a_{lm}|^2 \rangle= 36 \pi^2 \, \frac{(l+2)!}{(l-2)!}\,
\int {\rm d}k\, A^2(k) \left|F_l(k)\right|^2,
\ee
where
\be
\label{eq:Fl(k)}
F_l(k)= \frac{1}{\sqrt{k}} \int^{\tau_{\rm now}}_{\tau_{\rm LSS}} {\rm d}\tau
\, \frac{\partial\,\,\,}{\partial\tau}\left[ \frac{1}{3}
\frac{h^i_{\bf k}(\tau)}{h^i_{\bf k}(0)} \right]\,
\frac{j_l(k(\tau_{\rm now}-\tau))}{[k(\tau_{\rm now}-\tau)]^2}.
\ee
where $\tau_{\rm now}$ is the conformal time today,
$\tau_{\rm LSS}$ is the conformal time at last scattering.
We need to find the graviton mode function $h^i_{\bf k}(\tau)$.

Inflation gives rise to a spatially flat and perturbed
Friedmann-Robertson-Walker universe with the metric
\be
g_{\mu\nu}=R^2(\tau)\,[\eta_{\mu\nu}+h_{\mu\nu}],
\ee
where $\eta_{\mu\nu}=diag(1,-1,-1,-1)$, $h_{\mu\nu}$ is a
small perturbation, and $\tau$ is the conformal time.
The cosmic scale factor $R(\tau)$ is
\be
\label{eq:R(tau)}
R(\tau)=\left[\tau/\tau_0+ \sqrt{R_{\rm eq}}\,\right]^2-R_{\rm eq},
\ee
for a Universe with both matter and radiation. We have defined
$\tau_0\equiv 2H_0^{-1}\sqrt{1+R_{\rm eq}}$.
At matter-radiation equality, $R(\tau_{\rm eq})\equiv R_{\rm eq}=4.18\times
10^{-5}h^{-2}$, and $\tau_{\rm eq}/\tau_0=[\sqrt{2}-1] R_{\rm eq}^{1/2} $.
Today $R_{\rm now}=1$, $\tau_{\rm now} /\tau_0=\sqrt{1+R_{\rm eq}}-
\sqrt{R_{\rm eq}}$. At last scattering, $R_{\rm LSS}=1/(1+z_{\rm LSS})$,
$\tau_{\rm LSS}/\tau_0=\sqrt{R_{\rm LSS}+R_{\rm eq}}-\sqrt{R_{\rm eq}}$.

The gravity-wave perturbation satisfies the massless Klein-Gordon
equation
\be
\label{eq:hik}
\ddot{h}^i_{\bf k}+2\left[ \frac{\dot{R}}{R}\right] \dot{h}^i_{\bf k}
+k^2\, h^i_{\bf k}=0,
\ee
where $k^2={\bf k}\cdot{\bf k}$ and the overdots denote derivatives with
respect to $\tau$.

A gravity-wave mode with wavenumber ${\bf k}$ crosses inside the
horizon at $k\tau \sim 1$. Before it crosses inside the
horizon, $k\tau \ll 1$. Eq.(\ref{eq:hik}) gives us
$\dot{h}^i_{\bf k}(\tau)=0$ for $k\tau \ll 1$, i.e., the gravity-wave mode
is frozen before horizon-crossing. We can take $\dot{h}^i_{\bf k}(0)=0$
as the initial condition for Eq.(\ref{eq:hik}). For modes that cross
inside the horizon during radiation dominated era ($\tau \ll \tau_{\rm eq}$,
$R(\tau)=2 \sqrt{R_{\rm eq} }\, \tau/\tau_0$),
the exact solution is $h^i_{\bf k}(\tau)=h^i_{\bf k}(0)\, j_0(k\tau)$;
for modes that cross inside the horizon during matter dominated era
($\tau \gg \tau_{\rm eq}$, $R(\tau)=(\tau/\tau_0)^2$)
the exact solution is $h^i_{\bf k}(\tau)=h^i_{\bf k}(0)\, 3j_1(k\tau)/(k\tau)$.
Here $j_0(z)=\sin z/z$ and $j_1(z)=\sin z/z^2-\cos z/z$ are spherical Bessel
functions of order zero and one respectively.

\section{3. First method: time-dependent transfer function}
Since the contributions to the tensor multipole moments are dominated
by gravity waves which have entered the horizon recently [\ref{White}],
let us write [\ref{TWL}]
\be
\label{eq:def T}
h_{\bf k}^i(\tau)=h_{\bf k}^i(0) T_{\tau}(k/k_{\rm eq}) \left[
\frac{3 j_1(k\tau)}{k\tau} \right],
\ee
where $T_{\tau}(k/k_{\rm eq})$ is the amplitude transfer function
which accounts for the effect of short-wavelength modes entering the
horizon during radiation-dominated era, and $k_{\rm eq} \equiv
\tau_{\rm eq}^{-1}$. Eq.(\ref{eq:Fl(k)}) becomes
\be
\label{eq:F_l(k)}
F_l(k)=-k^{3/2}\, \int^{\tau_{\rm now}}_{\tau_{\rm LSS}}{\rm d}\tau\,
\tau \,T_{\tau}(k)\,\frac{j_2(k\tau)}{(k\tau)^2}
\frac{j_l(k(\tau_{\rm now}-\tau))}{[k(\tau_{\rm now}-\tau)]^2}.
\ee
Note that there is no $\partial T_{\tau}(k)/\partial\tau$ term in
the above expression, because $\partial h^i_{\bf k}(\tau)/\partial\tau$
and $h^i_{\bf k}(\tau)$ are related to $3j_1(k\tau)/(k\tau)$ and
$\partial[3j_1(k\tau)/(k\tau)]/\partial\tau$ by
the same amplitude transfer function $T_{\tau}(k)$.

The transfer function at time $\tau$ can be found by integrating
Eq.(\ref{eq:hik}) numerically from $\tau=0$ to $\tau$. Today's
transfer function is [\ref{TWL}]
\be
\label{eq:T_0(k)}
T_{\tau_0}(y) \equiv T_0(y)= \left[1.0+1.34y+2.5y^2\right]^{1/2},
\ee
where $y \equiv k/k_{\rm eq}$.

Since the Universe became
matter-dominated {\it gradually}, the transfer function in
Eq.(\ref{eq:def T}) should obviously depend on time. Once a mode is well inside
the horizon ($k\tau \gg 1$), $h_{\bf k}^i(\tau) \propto \cos(k\tau)/R$
(see Eq.(\ref{eq:hik})). Since $3 j_1(k\tau)/(k\tau)$ is the exact mode
function for $R(\tau)=(\tau/\tau_0)^2$, the transfer function
for modes with $k \gg k_{\rm eq}$ at an early time $\tau$ is given by
[\ref{TWL}]
\be
\label{eq:T_tau^0(y)}
T_{\tau}^0(k/k_{\rm eq})= \frac{(\tau/\tau_0)^2}{R(\tau)}\, T_0(k/k_{\rm eq})
\equiv A(\tau)\, T_0(k/k_{\rm eq}), \hskip 2cm k\gg k_{\rm eq}.
\ee
The above formula is in very good agreement with numerical results
for $k\ga k_{\rm eq}$ and $\tau_{\rm LSS} \leq \tau \leq \tau_0$.

On the other hand, modes with $k \ll k_{\rm eq}$ entered the horizon during
matter-dominated era; the transfer function for these modes should have
negligible time dependence.
Let us write
\be
T_{\tau}(k/k_{\rm eq})=T_0(k/k_{\rm eq})\, T_1(\tau,k/k_{\rm eq}),
\ee
where $T_0(k/k_{\rm eq})$ is given by Eq.(\ref{eq:T_0(k)}), and
$T_1(\tau,k/k_{\rm eq})$ can be written as
\be
T_1(\tau,k/k_{\rm eq})=w(k/k_{\rm eq}) A(\tau)+[1-w(k/k_{\rm eq})].
\ee
where $A(\tau) \equiv (\tau/\tau_0)^2/R(\tau)$.
$w(k/k_{\rm eq}) \rightarrow 0$ for $k \ll k_{\rm eq}$, and
$w(k/k_{\rm eq}) \rightarrow 1$ for $k \gg k_{\rm eq}$.
The simplest choice is
\be
\label{eq:w1}
w(k/k_{\rm eq})=1-\exp\left[-\gamma\, (k/k_{\rm eq})^{\Delta} \right],
\ee
where $\gamma$ and $\Delta$ are constants.

Since we use the graviton mode function from matter-dominated era,
it is consistent to use $\tau_{\rm LSS}\simeq \tau_0 \sqrt{R_{\rm LSS}}$
(which is the matter-dominated limit of the correct expression)
as the lower limit of integration in Eq.(\ref{eq:Fl(k)}).
With $\gamma=0.9$ and $\Delta=0.45$ in Eq.(\ref{eq:w1}),
the multipole moments computed using our transfer function agrees
reasonably well with the ``Boltzmann'' result for $l\la 50$ (see Fig. 1).
To get better result at larger $l$, we can use a slightly complicated weight
function
\be
\label{eq:w2}
w(k/k_{\rm eq})=\left(1-\exp\left[-\gamma\, (k/k_{\rm eq})^{\Delta} \right]
\right)\, \left(1-\exp\left[-n_c (k/k_{\rm eq}-y_c)^2 \right]\right),
\ee
where $n_c$ and $y_c$ are constants. With
$\gamma=0.9$, $\Delta=0.45$, $n_c=20$, $y_c=0.7$,
the multipole moments computed using our transfer function agrees reasonably
well with the ``Boltzmann'' result for $l\la 100$.
In Figure 1, we plot the CBR anisotropy tensor multipole spectrum
computed using the TWL transfer function (dotted line), the
transfer functions with weight functions given by Eq.(\ref{eq:w1})
(dashed line), and Eq.(\ref{eq:w2}) (dot-dashed line).
The solid line is the ``Boltzmann'' result.

It can be argued that using the matter-dominated limit for $\tau_{\rm LSS}$
distorts the ionization history of the Universe, since
the correct expression $\tau_{\rm LSS}=\tau_0\left[\sqrt{R_{\rm LSS}+
R_{\rm eq}}-\sqrt{R_{\rm eq}} \right]$ is smaller than
the matter-dominated limit $\tau_{\rm LSS}\simeq \tau_0 \sqrt{R_{\rm LSS}}$
by a factor of $2/3$. However, using the correct expression for
$\tau_{\rm LSS}$ increases the multipole moments for an amount which
increases from 10\% at $l=20$ to 300\% at $l=100$. The reason for this
dramatic effect is that the graviton mode function we use is an extremely
bad approximation to the true mode function at the era of last scattering.
The contribution to the multipole moment for a given $l$ is dominated by
the wavenumber at which $F_l(k)$ (see Eq.(\ref{eq:Fl(k)})) peaks;
since $F_l(k)$ peaks at $k\tau_0 \sim l$, i.e, $k/k_{\rm eq}
\sim 2.678 h^{-1} \, l\times 10^{-3}$ [\ref{TWL}], larger $l$ multipole moments
are dominated by the contribution from larger-wavenumber graviton modes,
which entered the horizon at earlier times, when the true graviton mode
function deviates greatly from the matter-dominated graviton mode function
that we use. Our transfer function can {\it not} correct for this effect even
with time dependence included, because at early times (around the
last scattering) the graviton mode function has only a smaller number
of oscillations in $k$, while our transfer function only accounts for
the difference in {\it average amplitude} between the true and the
matter-dominated
mode functions.

By using the matter-dominated limit for $\tau_{\rm LSS}$,
we are effectively truncating the integral over conformal time $\tau$ in the
expression for the multipole moment; it is not surprising that this enables us
to get multipole moments (for $l \la 120$)
which are not far off from the ``Boltzmann'' results, since
we are cutting off the $\tau$ integral at small $\tau$ where the
matter-dominated mode function is most inaccurate.
The multipole moments we obtain by using
the matter-dominated limit for $\tau_{\rm LSS}$ are therefore physically
consistent.

The transfer function method is limited to $l \la 120$. For larger $l$,
the phase difference between the matter-dominated graviton mode function
and the true mode function becomes important, which results in significant
deviation between the multipole spectrum computed using
{\it any} transfer function
and the multipole spectrum from the ``Boltzmann'' method.

\section{4. Second method: approximate gravity-wave mode function}
If we want to use the correct expression for $\tau_{\rm LSS}$,
we $must$ use a new mode function which better approximates the true
graviton mode function at small $\tau$ than the matter dominated
mode function $3j_1(k\tau)/(k\tau)$.

Obviously, one can construct an approximate solution which interpolates
smoothly between $j_0(k\tau)$ for $\tau \ll \tau_{\rm eq}$ and
$3j_1(k\tau)/(k\tau)$ for $\tau \gg \tau_{\rm eq}$.
Let us write
\ba
\label{eq:hnt}
\frac{\partial\,\,}{\partial \tau}
\left[\frac{h_{\bf k}^i(\tau)}{h_{\bf k}^i(0)}\right]&=&
[1-w(\tau)] \, T_{\tau}^0(k/k_{\rm eq}) \,
\frac{\partial\,\,}{\partial \tau}\left[\frac{3 j_1(k\tau)}{k\tau} \right]
+w(\tau)\,\frac{\partial j_0(k\tau)}{\partial \tau},\nonumber\\
&=& -k\, \left\{
[1-w(\tau)]\, T_{\tau}^0(k/k_{\rm eq}) \left[\frac{3 j_2(k\tau)}{k\tau} \right]
+w(\tau)\, j_1(k\tau)\right\},
\ea
where $w(\tau) \rightarrow 0$ for $\tau \gg \tau_{\rm eq}$,
and $w(\tau) \rightarrow 1$ for $\tau \ll \tau_{\rm eq}$.
$T_{\tau}^0(k/k_{\rm eq})= [(\tau/\tau_0)^2/R(\tau)]\, T_0(k/k_{\rm eq})$
(see Eq.(\ref{eq:T_tau^0(y)})), it correctly accounts for the difference
in average amplitude between the matter-dominated mode function and the true
mode function.

Since $j_0(k\tau)$ accurately gives both the amplitude and the phase of
the true mode function at large $k$, we should ``turn off''
$T_{\tau_0}(k/k_{\rm eq})$ (which accounts for the amplitude difference
between $3j_1(k\tau)/(k\tau)$ and the true mode function) at large $k$.
We can replace Eq.(\ref{eq:T_0(k)}) with
\be
\label{eq:T_0(k)newh}
T_0(y)=e^{-a y^b}\, \left[1.0+1.34y+2.5y^2\right]^{1/2}+ \left(1-e^{-a y^b}
\right), \hskip 1cm y\equiv k/k_{\rm eq}.
\ee

To compute the multipole moments, we make the following substitution
in Eq.(\ref{eq:F_l(k)}):
\be
T_{\tau}(k) \,\frac{ j_2(k\tau)}{k\tau} \Longrightarrow
[1-w(\tau)]\,T_{\tau}^0 \left(k/k_{\rm eq}\right) \,
\frac{ j_2(k\tau)}{k\tau}+w(\tau) \, \frac{j_1(k\tau)}{3}.
\ee
A good choice for $w(\tau)$ is
\be
\label{eq:w-newh}
w(\tau)=e^{-\alpha (\tau/\tau_{\rm eq})^\beta},
\ee
where $\alpha$ and $\beta$ are constants.

Fitting the resultant tensor multipole spectrum to that found by the
``Boltzmann'' method, we find $\alpha=0.2$, $\beta=0.65$, $a=b=4$ (the
fitting is more sensitive to $\alpha$ and $\beta$ than to $a$ and $b$).
The agreement between our result with the ``Boltzmann'' result is
impressive: better than 2\% for $l\la 50$, better than 10\% for $l\la 120$,
and better than 20\% for $ l \leq 300$. In Figure 2, we plot
the CBR anisotropy tensor multipole spectrum computed using
the matter-dominated graviton mode function with the TWL transfer function
(dotted line); our approximate graviton mode function (given by
Eqs.(\ref{eq:hnt}), (\ref{eq:T_0(k)newh}), and (\ref{eq:w-newh}))
(dashed line). The solid line is the ``Boltzmann'' result.

The multipole spectrum computed using the ``Boltzmann'' method
has a peak at $l\simeq 217$ (see Figure 2).
The multipole spectrum computed using our approximate mode
function has a peak at $l\simeq 213$, while the multipole spectrum
computed using the matter-dominated mode function
(with or without transfer function) has a peak at $l\simeq 180$ (see Figure 2).
Our approximate mode function gives much more accurate phase
information than the matter-dominated mode function.

The successful application of our approximate graviton mode function in
computing the tensor multipole spectrum stems from the fact that it is rather
close to the true graviton mode function.
In Figures 3 and 4, we plot the conformal time derivatives
of the true mode function (solid line), our approximate mode function
(given by Eqs.(\ref{eq:hnt}), (\ref{eq:T_0(k)newh}), and (\ref{eq:w-newh}))
(dashed line), and the matter dominated mode function (dotted line).
Figure 3 shows that at $\tau=\tau_{\rm LSS}$,
our approximate mode function is much closer to the true mode function
than the matter-dominated mode function, for all wavenumbers $k$.
Figure 4 shows that for $k=k_{\rm eq}\equiv \tau_{\rm eq}^{-1}$,
our approximate mode function is also much closer to
the true mode function than the matter-dominated mode function.

\section{5. Generalization}
The CBR anisotropy tensor multipole spectrums in Figures 1 and 2 are for
the standard values of $h=0.5$, $\Omega_{\rm B}=0.05$, $\Omega_0=1$;
and for a scale-invariant primordial spectrum of gravity waves.
It is straightforward to use our methods to compute the tensor multipole
spectrum for a non-scale-invariant primordial spectrum of gravity waves,
just use the corresponding $A(k)$ (see Eq.({\ref{eq:A(k)}),
which is constant in
the scale-invariant case) in Eq.(\ref{eq:|a_lm|^2}).

Next, we consider the cases when some or all of the parameters $h$,
$\Omega_{\rm B}$, and $\Omega_0$ are different from the standard values
($h=0.5$, $\Omega_{\rm B}=0.05$, $\Omega_0=1$).
$z_{\rm LSS}$ is given by [\ref{KT}]
\be
1+z_{\rm LSS} \simeq 1100\, (\Omega_0/\Omega_{\rm B})^{0.018}.
\ee
As long as $\Omega_0=1$, the cosmic scale factor is given by
Eq.(\ref{eq:R(tau)});
our previous formalism applies with $z_{\rm LSS}$ given above.
In Figure 5, we plot the CBR anisotropy
tensor multipole spectrum computed using our approximate graviton mode
function (given by Eqs.(\ref{eq:hnt}), (\ref{eq:T_0(k)newh}),
and (\ref{eq:w-newh})) for
$\Omega_0=1$, $\Omega_{\rm B}=0.05$, $h=0.8$ (solid line) and
$h=0.5$ (dashed line).

If $\Omega=\Omega_0+\Omega_\Lambda=1$, but $\Omega_0<1$, we have
\ba
&&\frac{\tau}{\tau_0}= \frac{1}{2} \int_0^{R(\tau)}\frac{ {\rm d}R}
{\sqrt{ R_{\rm eq}+R+(\Omega_{\Lambda} /\Omega_0)(1+R_{\rm eq})\, R^4}},
\nonumber\\
&& \tau_0 \equiv 2 \Omega_0^{-1/2} H_0^{-1} \sqrt{1+R_{\rm eq}},
\nonumber\\
&& R_{\rm eq}= 4.18 \times 10^{-5} (\Omega_0 h^2)^{-1},\nonumber\\
&& k_{\rm eq} \equiv \tau_{\rm eq}^{-1}=\frac{\tau_0^{-1}}
{(\sqrt{2}-1) \sqrt{R_{\rm eq}} }, \nonumber\\
&&  \frac{\tau_{\rm LSS}}{\tau_0}= \sqrt{R_{\rm LSS}+R_{\rm eq}}-
\sqrt{R_{\rm eq}}.
\ea
Note that the conformal time today $\tau_{\rm now} \neq \tau_0$.
For $h=0.8$, $\Omega_0=0.2$, $\Omega_\Lambda=0.8$,
$\tau_{\rm now} /\tau_0 \simeq 0.85$.
A sizable cosmological constant term significantly alters the late time
evolution of the cosmic scale factor, hence
$3j_1(k\tau)/(k\tau)$ is no longer the mode function at late times.
However, our approximate graviton mode function is still a good
approximation of the true mode
function in the context of computing tensor multipole spectra,
which are not very sensitive to the late time graviton mode
function [\ref{TWL}]. In Figure 6, we plot the CBR anisotropy
tensor multipole spectrum computed using our approximate graviton mode
function for $\Omega_0=0.2$, $\Omega_{\Lambda}=0.8$, $\Omega_{\rm B}=0.02$,
$h=0.8$ (solid line), and $\Omega_0=1$, $\Omega_{\rm B}=0.05$, $h=0.5$
(dashed line).

We expect the tensor multipole spectrum computed using our approximate
graviton mode function (given by Eqs.(\ref{eq:hnt}), (\ref{eq:T_0(k)newh}),
and (\ref{eq:w-newh}), with $\alpha=0.2$, $\beta=0.65$, $a=b=4$)
to have fairly good accuracy when the cosmological parameters $h$,
$\Omega_{\rm B}$, and $\Omega_0$ differ from the standard values
of $h=0.5$, $\Omega_{\rm B}=0.05$, $\Omega_0=1$.

\section{6. Discussion}

We have presented two simple and straightforward
analytical methods for computing the CBR anisotropy tensor multipole spectrum;
one method uses a time-dependent transfer function,
the other method uses an approximate gravity-wave
mode function which is a simple combination of the lowest order
spherical Bessel functions (given by Eqs.(\ref{eq:hnt}),
(\ref{eq:T_0(k)newh}), and (\ref{eq:w-newh}), with $\alpha=0.2$,
$\beta=0.65$, $a=b=4$). Both methods give much better accuracy
than using the matter-dominated mode function with the
time independent transfer function of Turner, White, and Lidsey [\ref{TWL}].
Our approximate graviton mode function method is especially
promising, since it gives a tensor multipole spectrum which is impressively
accurate (difference with the ``Boltzmann result'' is less than 2\% for
$l \la 50$, less than 10\% for $l \la 120$, and less than 20\% for $l\leq
300$).

Many attempts have been made in the past in finding a good
approximate graviton mode function, for instance, via the
sudden approximation (assuming the transition from radiation-domination
to matter-domination to be instantaneous) [\ref{sudden}].
In addition, Ng and Speliotopoulos explored the possibility of finding
a good approximate graviton mode function via the WKB approximation
[\ref{NS}]; however, they did not use the mode function they found
in computing the multipole moments, they used
the mode function found by numerical integration instead.
Furthermore, Koranda and Allen found the exact graviton mode function in
terms of spheroidal wavefunctions [\ref{KA}]. The advantage of our approximate
graviton mode function is that it takes into account
the {\it gradual} transition from radiation-domination to matter-domination
in the Universe; it can be used to compute the CBR anisotropy tensor
multipole spectrum rather accurately;
and it involves only the first and second order spherical Bessel functions.

In a different direction of the CBR anisotropy tensor multipole spectrum
calculation, the ``Boltzmann method'' evolves the photon distribution function
numerically, using first-order perturbation theory of the general relativistic
Boltzmann equation for radiative transfer [\ref{DKK}].
We have used the CBR anisotropy tensor multipole spectrum from the
``Boltzmann method'' as the standard for comparison, but we do not think
that it is practical to use the numerically rather involved
``Boltzmann method'' in all cases. When we study
the tensor perturbations from a large number of inflationary models,
it is much more convenient to use the Sachs-Wolfe formula with a simple
yet reasonably accurate graviton mode function, such as the one presented in
this paper.


\vskip 0.2in
\centerline{\bf Acknowledgments}
I am indebted to Scott Dodelson and Lloyd Knox for their generosity of
providing me with their data of the ``Boltzmann method''. I thank
Michael Turner, Rocky Kolb, Albert Stebbins, and Scott Dodelson for very
helpful discussions. This work was supported by the DOE and NASA under
Grant NAG5-2788.

\newpage
\frenchspacing
\parindent=0pt

\centerline{{\bf References}}

\begin{enumerate}

\item\label{SW} R.K. Sachs and A.M. Wolfe, Astrophys. J. {\bf 147}, 73 (1967).

\item\label{DKK} S. Dodelson, L. Knox, and E.W. Kolb, Phys. Rev. Lett.
{\bf 72}, 3444 (1994); S. Dodelson and L. Knox, private communication;
R. Crittenden, J.R. Bond, R.L. Davis, G. Efstathiou, and
P.J. Steinhardt, Phys. Rev. Lett. {\bf 71}, 324 (1993).

\item\label{KA} S. Koranda and B. Allen, WISC-MILW-94-TH-18.

\item\label{TWL} M.S. Turner, M. White, and J.E. Lidsey, Phys. Rev.
D{\bf 48}, 4613 (1993).

\item\label{KT} E.W. Kolb and M.S. Turner, {\it The Early Universe}
(Addison-Wesley, Redwood City, CA, 1990), and references therein.

\item\label{White} A.A. Starobinskii, Pis'ma Astron. Zh. {\bf 11}, 323 (1985)
[Sov. Astron. Lett. {\bf 11}, 133 (1985)]; L. Abbott and M. Wise,
Nucl. Phys. {\bf B244}, 541 (1984);
M. White, Phys. Rev. D {\bf 46}, 4198 (1992);
R. Fabbri, F. Lucchin, and S. Matarrese, Phys. Lett. {\bf 166B}, 49 (1986).

\item\label{sudden} A.G. Doroshkevich, I.D. Novikov, and A.G. Polnarev,
Astron. Zh. {\bf 54}, 932 (1977) [Sov. Astron. {\bf 21}(5), 529 (1977)];
L.F. Abbott and D.D. Harari, Nucl. Phys. {\bf B264}, 487 (1986);
L.P. Grishchuk, Phys. Rev. {\bf D48}, 3513 (1993);
K.-W. Ng, IP-ASTP-31-1993.

\item\label{NS} K.-W. Ng and A.D. Speliotopoulos, IP-ASTP-07-94.

\end{enumerate}

\newpage
\nonfrenchspacing
\parindent=20pt

\centerline{{\bf Figure Captions}}

\vspace{0.2in}

Figure 1. The CBR anisotropy tensor multipole spectrum computed using
various transfer functions, compared to the Boltzmann result (solid line).

Figure 2. The CBR anisotropy tensor multipole spectrum computed using
our approximate graviton mode function (given by Eqs.(\ref{eq:hnt}),
(\ref{eq:T_0(k)newh}), and (\ref{eq:w-newh})), compared to the Boltzmann
result (solid line).

Figure 3. The conformal-time derivatives at $\tau=\tau_{\rm LSS}$, of
the true graviton mode function (solid line), our approximate mode function
(given by Eqs.(\ref{eq:hnt}), (\ref{eq:T_0(k)newh}), and (\ref{eq:w-newh}))
(dashed line), and the matter-dominated mode function (dotted line).

Figure 4. The conformal-time derivatives
with $k=k_{\rm eq}\equiv \tau_{\rm eq}^{-1}$, of
the true graviton mode function (solid line), our approximate mode function
(given by Eqs.(\ref{eq:hnt}), (\ref{eq:T_0(k)newh}), and (\ref{eq:w-newh}))
(dashed line), and the matter-dominated mode function (dotted line).

Figure 5. The CBR anisotropy
tensor multipole spectrum computed using our approximate graviton mode
function (given by Eqs.(\ref{eq:hnt}), (\ref{eq:T_0(k)newh}), and
(\ref{eq:w-newh})) for $\Omega_0=1$, $\Omega_{\rm B}=0.05$, $h=0.8$
(solid line) and $h=0.5$ (dashed line).

Figure 6. The CBR anisotropy
tensor multipole spectrum computed using our approximate graviton mode
function (given by Eqs.(\ref{eq:hnt}), (\ref{eq:T_0(k)newh}), and
(\ref{eq:w-newh})) for
$\Omega_0=0.2$, $\Omega_{\Lambda}=0.8$, $\Omega_{\rm B}=0.02$, $h=0.8$
(solid line) and $\Omega_0=1$, $\Omega_{\rm B}=0.05$, $h=0.5$ (dashed line).

\end{document}